\documentclass[aps,prb,reprint,showpacs,superscriptaddress]{revtex4-1}
\usepackage{hyperref}
\usepackage{graphicx}
\usepackage{amsmath}
\usepackage{amssymb}
\usepackage{tipa}
\usepackage[caption=false]{subfig}
\usepackage{xcolor}
\usepackage{mathrsfs}
\usepackage{float}
\usepackage{tikz}
\usetikzlibrary{arrows}
\usetikzlibrary{shapes.geometric}

\definecolor{light-gray}{gray}{0.85}

\renewcommand{\tilde}{\widetilde}
\renewcommand{\hat}{\widehat}
\renewcommand{\leq}{\leqslant}
\renewcommand{\geq}{\geqslant}

\newcommand{\Tr}{\operatorname{Tr}}
\newcommand{\nn}{\nonumber}

\newcommand{\SU}{\operatorname{SU}}

\newcommand{\ZZ}{\mathbb{Z}}

\newcommand{\calH}{\mathcal{H}}

\begin{document}

\title{Holographic duality between $(2+1)$-d quantum anomalous Hall state and $(3+1)$-d topological insulators}

\author{Yingfei Gu}
\affiliation{Department of Physics, Stanford University, Stanford, CA 94305, USA}
\author{Ching Hua Lee}
\affiliation{Institute of High Performance Computing, 138632, Singapore}
\author{Xueda Wen}
\affiliation{Institute for Condensed Matter Theory and Department of Physics, University of Illinois at Urbana-Champaign, 1110 West Green St, Urbana IL 61801}
\author{Gil Young Cho}
\affiliation{Institute for Condensed Matter Theory and Department of Physics, University of Illinois at Urbana-Champaign, 1110 West Green St, Urbana IL 61801}
\affiliation{Department of Physics, Korea Advanced Institute of Science and Technology, Daejeon 305-701, Korea}
\author{Shinsei Ryu}
\affiliation{Institute for Condensed Matter Theory and Department of Physics, University of Illinois at Urbana-Champaign, 1110 West Green St, Urbana IL 61801}
\author{Xiao-Liang Qi}
\affiliation{Department of Physics, Stanford University, Stanford, CA 94305, USA}
\date{\small \today}

\begin{abstract}
In this paper, we study $(2+1)$-dimensional quantum anomalous Hall states, i.e. band insulators with quantized Hall conductance, using the exact holographic mapping. The exact holographic mapping is an approach to holographic duality which maps the quantum anomalous Hall state to a different state living in $(3+1)$-dimensional hyperbolic space. By studying topological response properties and the entanglement spectrum, we demonstrate that the holographic dual theory of a quantum anomalous Hall state is a $(3+1)$-dimensional topological insulator. The dual description enables a new characterization of topological properties of a system by the quantum entanglement between degrees of freedom at different length scales.
\end{abstract}

\maketitle

\section{Introduction}

Holographic duality, also known as the anti-de-Sitter space (AdS)/conformal field theory (CFT) duality, is a duality between a $(d+1)$-dimensional \textit{gravity theory} (bulk theory) and a $d$-dimensional quantum field theory\cite{maldacena1997large,witten1998anti,witten1998anti2,gubser1998gauge} (boundary theory). Originally proposed for the $(3+1)$-dimensional $\SU(N)$ super Yang-Mills theory, holographic duality has been generalized to many other systems. From the point of view of a $d$-dimensional quantum field theory, the additional dimension in the dual $(d+1)$-dimensional theory can be interpreted as an energy scale (or length scale), such that the dual theory is a generalization of renormalization group approach to the quantum field theory\cite{de2000holographic,freedman1999renormalization}. With such a physical interpretation, holographic duality has been applied as a new way to describe strongly correlated quantum many-body systems
\cite{hartnoll2009lectures,mcgreevy2010holographic}.

Although there has been a lot of evidence for holographic duality in various theories, a microscopic description of this duality is in general an open question. More specifically, for a given quantum field theory it is generally unknown whether the dual theory exists and, if it exists, how is the dual theory described. As an effort towards a microscopic description, an approach named as exact holographic mapping (EHM) has been proposed\cite{qi2013exact} by one of us. Inspired by the role of quantum entanglement in holographic duality\cite{ryu2006holographic,swingle2012entanglement,vidal2008class,haegeman2013entanglement,nozaki2012holographic}, the EHM approach defines a unitary mapping from the boundary theory to the bulk theory, and proposes a measure of the bulk geometry (for the coordinate choice given by the mapping) determined by the bulk correlation functions. More details about this approach will be discussed later in this article.

In this work, we apply the EHM approach to a topological state of matter, the quantum anomalous Hall insulator (also known as the Chern insulator). Topological states of matter are gapped states which are distinct from trivial insulators by topological properties, such as robust gapless edge states, topological response properties and/or ground state degeneracy. The integer quantum Hall state is the first electronic topological state of matter \cite{klitzing1980new}. In the past decade, the understanding to topological states of matter has been greatly expanded since the discovery of time reversal invariant topological insulators and topological superconductors\cite{qi2010quantum,qi2011topological,hasan2010colloquium,moore2010birth}. Naively, one may expect the holographic dual theory of a topological state to be trivial, since there is no low energy excitations in the infrared (IR) limit. Usually, the dual geometry of such a gapped system terminates at a ``end-of-the-world brane" at the infrared end, with its position in the emergent direction determined by the energy gap. For topologically ordered states with a fixed point wavefunction description\cite{levin2005string}, the multiscale entanglement renormalization ansatz (MERA) approach has been applied to construct a dual theory that has the topological information completely encoded at the infrared brane\cite{aguado2008entanglement}. 
\footnote{
In a companion paper \onlinecite{WenMERA}, the continuous MERA construction of the ground states of fermionic topological insulators (Chern insulators) in (2+1)-d, together with its emergent bulk geometry,
is discussed.}
However, as we will show in this work, the situation is more nontrivial for fermionic topological states such as quantum Hall states, since a fixed point wavefunction with zero correlation length does not exist.

\begin{figure}[h]
\center
\includegraphics[scale=0.35]{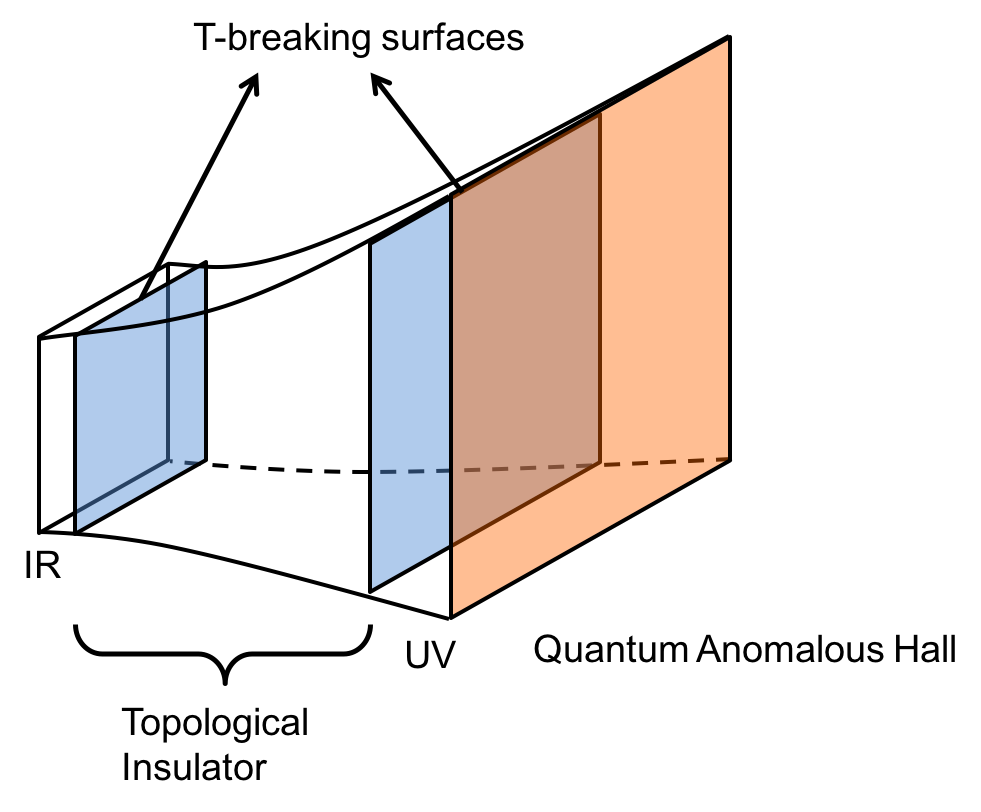}
\caption{Illustration of the duality between a topological insulator and a quantum anomalous Hall system.}
\label{Holographic TI}
\end{figure}

By applying the EHM approach to a quantum anomalous Hall (QAH) state, i.e. a quantum Hall state without an orbital magnetic field\cite{haldane1988model,qi2006topological,liu2008quantum,yu2010quantized,chang2013experimental}, we show that the dual bulk theory is a three-dimensional time reversal invariant topological insulator (TI), see Fig.~\ref{Holographic TI}. The topological insulator occupies a slab of finite thickness in the emergent direction, and the time reversal breaking necessary for the original quantum Hall state occurs at its surfaces. Physically, the position of the bulk TI surfaces correspond to length scales which contribute nontrivially to the topological invariant, i.e. the quantized Hall conductance. The fact that the bulk is a TI is a direct manifestation of the nontrivial quantum entanglement between degrees of freedom at different length scales in a QAH state. As an explicit verification of the bulk topological property, we study the entanglement spectrum of the system with an entanglement cut in the emergent bulk direction. This corresponds to the study of the entanglement between long-wavelength and short-wavelength degrees of freedom in the QAH state, which are well-defined only after performing the EHM transformation. With such a bulk entanglement cut, we obtain a gapless entanglement spectrum with odd number of massless Dirac cones, a fact that directly verifies the TI nature of the bulk theory\cite{li2008entanglement,fidkowski2010entanglement,turner2010entanglement,qi2012general}. Our result can be generalized straightforwardly to free fermion topological states in higher dimensions or in other symmetry classes, which suggests that nontrivial quantum entanglement between different energy scales and the duality between states in different dimensions are generic features of fermionic topological states of matter.

The remainder of this article will be organized as follows. In section \ref{sec:review}, we briefly review the exact holographic mapping, and generalize it to $(2+1)$-d fermion systems. In section \ref{sec:bulktheory1} we study the bulk dual theory of a $(2+1)$-d QAH state, and demonstrate its topological nature by a Chern number density calculation. In section \ref{sec:bulktheory2} we study the entanglement spectrum of the bulk theory, which provides further demonstration of its TI nature, and also defines a new entanglement probe of the QAH state. Finally, the conclusion and discussions are given in section \ref{sec:conclusion}.

\section{Brief review of the exact holographic mapping(EHM)}\label{sec:review}

\begin{figure}[htb]
\includegraphics[scale=0.34]{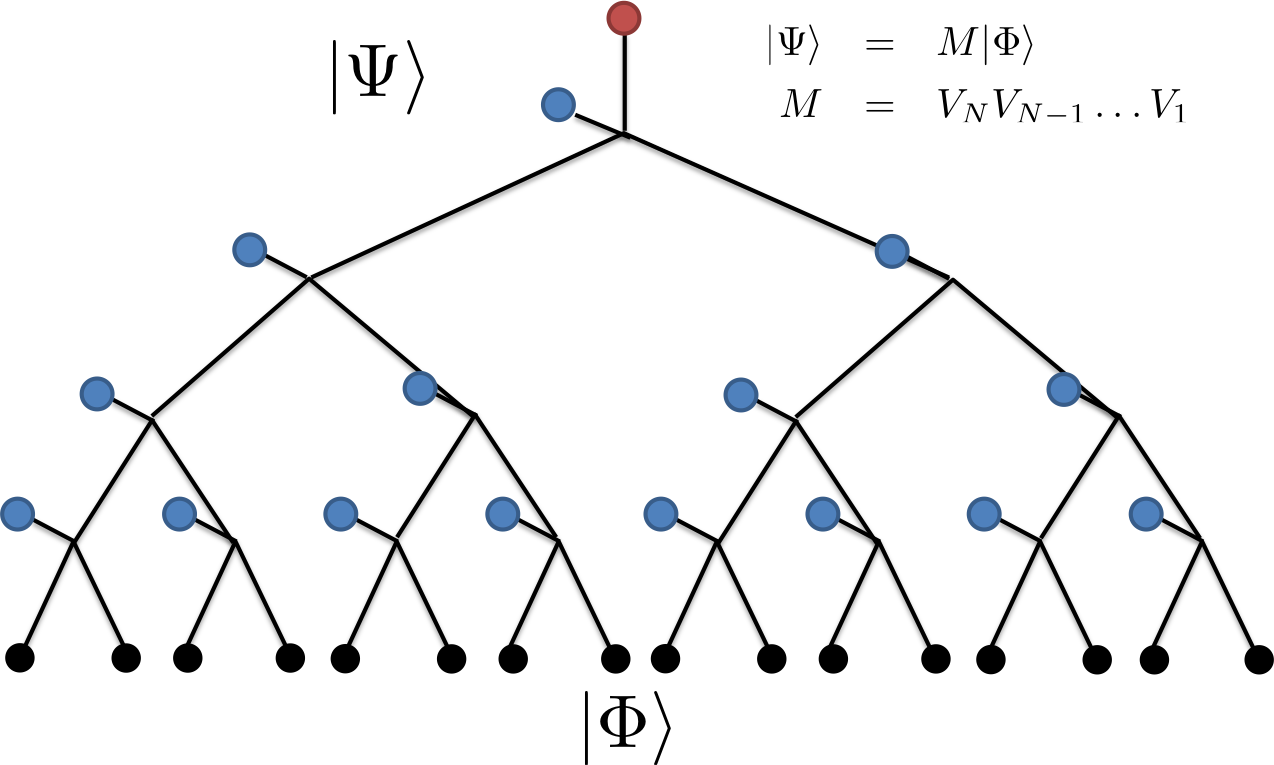}
\caption{A (simplest) example of the tensor network used to define the EHM. Each vertex is a two-site unitary transformation  that transforms (in the upward direction) two input sites (below) to two output sites (above). One of the output sites (blue dot) is considered as a bulk site, which contains the high energy degrees of freedom of the two input sites. The other site, containing the low energy degrees of freedom, becomes the input of the next layer (level in the tree). The whole network is a composition of these layers of unitary operators, which defines the unitary mapping $M$. The number of bulk sites (blue dots and the red dot at the top) is equal to the number of boundary sites (black dots). Bulk sites closer (farther) from boundary correspond to higher energy (lower energy) degrees of freedom.}
\label{fig:EHM}
\end{figure}

Ref.~\onlinecite{qi2013exact} proposed a new approach to holographic duality known as the exact holographic mapping (EHM). An EHM is a unitary mapping $M: \calH_1\rightarrow \calH_2$ that maps between two Hilbert spaces $\calH_1$ and $\calH_2$. Since the mapping is unitary, the two Hilbert spaces have the same dimension. However, the mapping $M$ is defined by a unitary tensor network (also known as a quantum circuit), such as the one in Fig.~\ref{fig:EHM}, which defines a tensor-product decomposition of $\calH_2$ into bulk ``lattice sites" that is different from the decomposition of $\calH_1$ into boundary sites. The emergent direction perpendicular to the boundary has physical meaning of energy scale.

Once the unitary mapping is defined, bulk correlation functions can be used to probe the geometry in the bulk, such that the geometry formed by bulk sites is ``emergent" from the correlation structure rather than given externally. More explicitly, the spatial geometry can be determined by computing the two-site mutual information $I(x,y)=S_x+S_y-S_{xy}$ between two bulk sites. Here $S_x$ and $S_y$ are the von Neumman entropy of two bulk sites (for a given state $\rho_{\rm bulk}=M\rho_{\rm boundary}M^\dagger$), and $S_{xy}$ is the joint entropy of the two sites together. The distance between these two sites is {\it defined} as
\begin{eqnarray}
d(x,y)=-\xi \log \frac{I(x,y)}{I_0}\label{distancedef}
\end{eqnarray}
Here $I_0=2\log D$, with $D$ the dimension of Hilbert space on each site, is the maximally possible two-site mutual information. $\xi$ is a correlation length, which is an overall unit of scale.

With a given definition of unitary mapping $M$ and the distance definition (\ref{distancedef}), the emergent geometry in the bulk can be studied for different boundary states. Refs.~\onlinecite{qi2013exact} and Ref.~\onlinecite{lee2015} studied the $(1+1)$-d free fermion model and investigated the bulk geometry corresponding to different states on the boundary such as massless fermion ground state, massive fermion ground state and massless fermion finite temperature ensemble, etc. In the examples therein, the two-site unitary mapping at each vertex of the EHM network is defined by the following single-particle unitary transformation:
\begin{eqnarray}
\left( \begin{tabular}{c}
$a_{i,n}$ \\
$b_{i,n}$
\end{tabular}
\right) = \frac{1}{
\sqrt{2}}
\left(
\begin{tabular}{c c}
$1$ & $1$ \\
$1$ & $-1$
\end{tabular}
\right)
\left( \begin{tabular}{c}
$a_{2i-1,n-1}$ \\
$a_{2i,n-1}$
\end{tabular}
\right)
\end{eqnarray}
When the transformation acts on single-particles, the EHM is mathematically equivalent to a wavelet transformation\cite{lee2015}. The particular choice of mapping here corresponds to Haar wavelets\cite{haar1910theorie,wilson1971renormalization,fishman2015}.

Composing the unitary mappings of each site, we have a mapping from a boundary $(1+1)$-d chain with $2^N$ sites labeled by annihilation operators $\lbrace a_{i,0}| 1\leq i \leq 2^N \rbrace$ to a bulk $(2+1)$-d lattice model defined on a network with sites number $2^N=2^{N-1}+2^{N-2}+\ldots+2+1+1$, with the annihilation operators on each site labeled by $\lbrace b_{i,1}| 1\leq i \leq 2^{N-1} \rbrace \cup \lbrace b_{i,2}| 1\leq i \leq 2^{N-2} \rbrace \cup \ldots \cup \lbrace b_{1,N}\rbrace \cup \lbrace a_{1,N}\rbrace$. For example, we can choose a boundary with two component spinors $a_{i,0}=c_i, 1\leq j \leq 2^N$ and a Dirac Hamiltonian in momentum space:
\begin{eqnarray}
\widehat{H}=\sum_k c^\dagger_k \left[ \sigma_x  \sin k + \sigma_z (m+1-\cos k) \right] c_k
\end{eqnarray}
Correspondingly all the transformed fermions $b_{i,n}$ and $a_{i,n}$ are also two-component spinors.

Following the definition of distance in the bulk (Eq. \ref{distancedef}), one can study the emergent space-time geometry for a given state. For the above Dirac model at mass $m=0$ and temperature $T=0$, the bulk geometry in large scale is scale-invariant and reduces to approximately the Anti-de Sitter space $AdS_3$. With $m=0$ and a finite temperature $T>0$, the geometry becomes a black hole geometry, which can be verified by both spatial correlation functions and Euclidean time-direction correlation functions. For the model with a finite mass $m\neq 0$ at zero temperature $T=0$, the geometry contains a spatial termination surface in infrared limit, while the red shift in time-direction remains finite at this surface. In this case the geometry is similar to a ``confinement geometry"\cite{witten1998anti}.

Although the original EHM for free fermion is defined for a $(1+1)$-d boundary system, it is straightforward to extend it to arbitrary dimensions\cite{lee2015} by declaring that at each step, we apply the transformation $V_i$ to all the directions.

\begin{figure}[htb]
\includegraphics[scale=0.2]{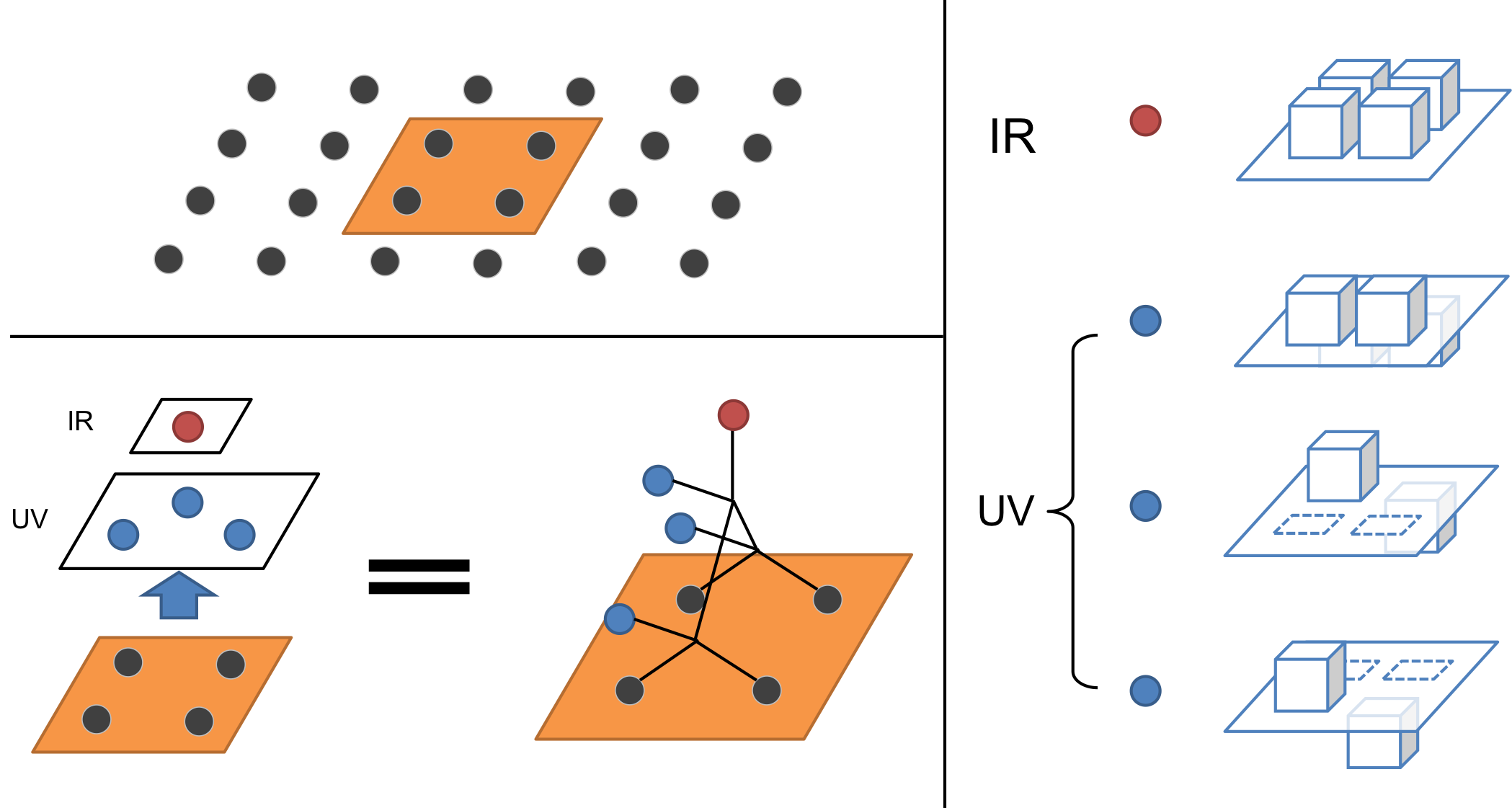}
\caption{(Left panel) An illustration of a transformation $V_n$ in two dimensional EHM (see Eq.~\ref{eqn: real space EHM}). The mapping is constructed by combining two $(1+1)$-d EHM steps, one in the $x$ direction and the other in the $y$ direction. The blue dots are the output UV modes and the red dot is the IR mode at present level, which will become the input for the next level. (Right panel) An illustration of the wavefunctions of the IR and UV modes, The IR mode is a uniform superposition of the 4 modes involved, while the UV modes are the three orthogonal oscillatory suppositions.
}
\label{fig:2+1d EHM}
\end{figure}
For instance, in $(2+1)$-d we can defined an EHM via the following two-step transformation (see Fig.~\ref{fig:2+1d EHM}):
\begin{eqnarray}
V_n&=&V^y_n V^x_n
\label{eqn: real space EHM}
\\
V^x_n:
\left( \begin{tabular}{c}
$\tilde{a}_{i,j,n}$ \\
$b^x_{i,j,n}$
\end{tabular}
\right) &=& \frac{1}{
\sqrt{2}}
\left(
\begin{tabular}{c c}
$1$ & $1$ \\
$1$ & $-1$
\end{tabular}
\right)
\left( \begin{tabular}{c}
$a_{2i-1,j,n-1}$ \\
$a_{2i,j,n-1}$
\end{tabular}
\right) \nonumber \\
V^y_n:
\left( \begin{tabular}{c}
$a_{i,j,n}$ \\
$b^{y}_{i,j,n}$
\end{tabular}
\right) &=& \frac{1}{
\sqrt{2}}
\left(
\begin{tabular}{c c}
$1$ & $1$ \\
$1$ & $-1$
\end{tabular}
\right)
\left( \begin{tabular}{c}
$\tilde{a}_{i,2j-1,n}$ \\
$\tilde{a}_{i,2j,n}$
\end{tabular}
\right)  \nonumber
\end{eqnarray}
In this paper, we will apply this mapping to the following $(2+1)$-d free fermion system with the Dirac Hamiltonian:
\begin{eqnarray}
\widehat{H}&=&\sum_{{\bf k}=\left(k_x,k_y\right)} c^\dagger_{{\bf k}} \left[ \vec{d}({\bf k}) \cdot \vec{\sigma} \right] c_{\bf k}
\label{eqn:two band chern insulator model}
  \\
\vec{d}({\bf k})&=&\left( \sin k_x, \sin k_y,m+2-\cos k_x-\cos k_y\right)
\nn
\end{eqnarray}
This simple two-band model has topological phases with quantized Hall conductance\cite{qi2006topological} for the ranges of mass $m\in(-2,0)$ and $m\in(-4,-2)$. Such a quantum Hall phase without orbital magnetic field is known as quantum anomalous Hall (QAH) state or Chern insulator\cite{haldane1988model,qi2006topological}. In the following, we will study the dual geometry of this system and especially the manifestation of its topological properties in the bulk theory. We will also discuss more generic models with QAH phases.

\section{Identifying the bulk theory (I): Chern number density}\label{sec:bulktheory1}

After defining the EHM that maps a $(2+1)$-d system to a $(3+1)$-d system, it is natural to ask whether the topological properties of the $(2+1)$-d model have counterparts in the $(3+1)$-d dual theory. In this section and next section, we will show two different evidences that the dual theory of a gapped Dirac model is a $(3+1)$-d topological insulator with T-breaking surfaces. The first evidence discussed in this section is the {\it Chern number density distribution} in the bulk system. The Chern number density, as we will define soon, determines the Hall conductance of the bulk system. By showing that the quantized Hall conductance of the $(2+1)$-d Dirac model is carried by two regions in the bulk, each carrying half quantum of the Hall conductance, we can prove the bulk theory is a TI.

\subsection{Chern number density and EHM}

We start by reviewing the definition of Chern number in a band insulator. In general, the single-particle Hamiltonian of a theory with $N$ energy bands can be written as a $N\times N$ matrix $h({\bf k})$. Denote $| s {\bf k}\rangle, s=1,2,...,N$ as the eigenstates of $h({\bf k})$, such that $h({\bf k})=\sum_s \epsilon_s ({\bf k})|s{\bf k}\rangle\langle s{\bf k}|$. If there are $M$ occupied bands denoted by $\alpha=1,2,...,M\leq N$, the Hall conductance of the system is $\sigma_H=\frac{e^2}h c_1$ with $c_1$ the Chern number defined as
\begin{eqnarray}
c_1=\frac{1}{2\pi} \int_{\rm BZ} d^2{\bf k} \operatorname{Tr} \left({\bf  f }_{\mu\nu} \right) = \sum_{\alpha=1}^M \frac{1}{2\pi} \int_{\rm BZ} d^2{\bf k}\   f^{\alpha\alpha}_{\mu\nu}
\label{Cherndef}
\end{eqnarray}
where $\textbf{f}_{\mu\nu}=\partial_\mu\textbf{a}_\nu-\partial_\nu \textbf{a}_\mu + [\textbf{a}_\mu, \textbf{a}_\nu]$ and $\lbrace \textbf{a}_\mu \rbrace$ are $M$ by $M$ matrices, whose matrix elements $\operatorname{a}_\mu^{\alpha\beta}(\textbf{k})=i \langle \alpha{\bf k} | \partial_\mu | \beta{\bf k} \rangle$ are the (non-abelian) Berry connections of occupied bands. Since the first Chern number $c_1$ only depends on the trace of the Berry curvature $\textbf{f}_{\mu\nu}$, the expression (\ref{Cherndef}) applies for any choice of basis $|\alpha' {\bf k}\rangle,~\alpha' =1,2,...,M$ that spans the space of occupied states, i.e. each $\alpha'$ does not have to label an energy band by itself, but can instead be a superposition of the bands.

In a $(3+1)$-d band insulator, the Hall conductance $\sigma_{ij}$ is not necessarily quantized. However, if we consider a thin film of a $(3+1)$-d material, with translation symmetry in the $x,y$ directions and open boundary condition in the $z$ direction, the total Hall conductance $\sigma_{xy}$ (integrated over the thickness in the $z$ direction) is still quantized, since one can view the thin film as a two-dimensional system and the third dimension as an ``internal" degree of freedom. Labeling the states by $z$-directional real space position $n$ and $x,y$-directional momenta ${\bf k}=(k_x,k_y)$, we can use a basis $|ns{\bf k}\rangle$, with $s$ labeling the energy levels at the same $n{\bf k}$. This basis defines a projector to $n$-th layer $\operatorname{P}_n({\bf k})=\sum_s|ns{\bf k}\rangle \langle ns{\bf k}|$. Using this projector we can define the Chern number density
\begin{eqnarray}
c(n):=\frac{1}{2\pi} \int_{\rm BZ} d^2{\bf k}   {\rm Tr}\left({\bf f}_{\mu\nu}({\bf k}) \operatorname{P}_n({\bf k})\right)\label{Cherndensity}
\end{eqnarray}
which is the contribution of $n$-th layer to the Hall conductance: $\sigma_{xy}(n)=c(n)\frac{e^2}{h}$. By definition, the sum of $c(n)$ gives the quantized net Chern number $c_1=\sum_nc(n)$.

\begin{figure}[htb]
{\includegraphics[scale=1]{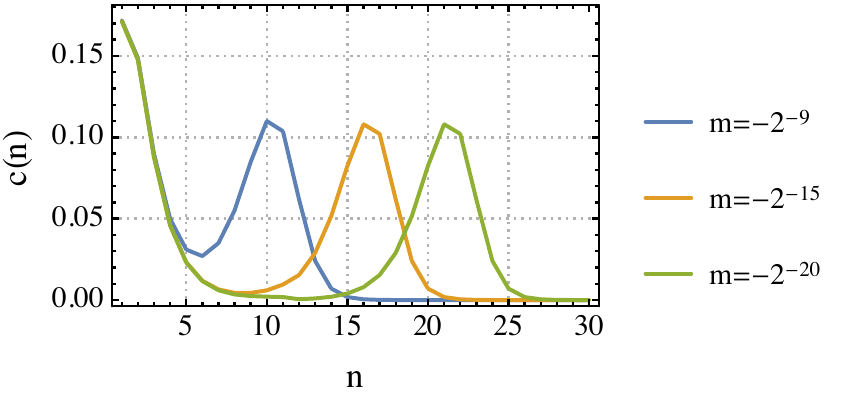}}
\caption{Numerical results on the Chern number density for Dirac model with different masses. We notice there are two peaks, the UV peak on the left and the IR peak on the right, each contributing a Chern number of approximately $\frac1{2}$. Generically, the IR peak arises from the Berry curvature contributions of the massive Dirac cone, while the UV peak arises from states near the Brillouin zone boundary, which provide UV regularization of the massive Dirac cone. 
}
\label{fig:cnumber_numerical}
\end{figure}

Now we study the Chern number density of the bulk obtained by EHM. Each step of EHM explicitly breaks lattice translation symmetry and doubles the unit cell in both directions. Denoting $T_x,T_y$ as the original lattice translation symmetries, after $N$ steps of mapping the residual translation symmetry is $T_x^{2^N}$ and $T_y^{2^N}$. We can always take a system size $\gg 2^N$ so that after $N$ steps of mapping the residual translation symmetry still acts nontrivially on the system. Therefore after $N$ steps the corresponding Brillouin zone will be folded into a $[-\frac{\pi}{2^N}, \frac{\pi}{2^N})\times [-\frac{\pi}{2^N}, \frac{\pi}{2^N})$ square. Each energy band in the original model is folded into $4^N$ bands in the reduced Brillouin zone. We can then explicitly write down the basis transformation in momentum space, see appendix~\ref{appendix: momentum space} for details and expressions. With the EHM basis transformation, it is straightforward to construct the projection operator and compute the Chern number density. The numerical data is shown in Fig.~\ref{fig:cnumber_numerical}.


\subsection{Identifying bulk holographic topological insulator}

The numerical results in Fig.~\ref{fig:cnumber_numerical} show two peaks in the Chern number density for the bulk EHM dual of the model~(Eq. \ref{eqn:two band chern insulator model}). The locations of the peaks corresponds to two different length scales that contribute to Berry curvature: the left peak at UV scale corresponding to the Berry curvature at high momentum around the Brillouin zone boundary, while the right peak corresponds to the Berry curvature accumulated near ${\bf k}=0$ (IR) point, which we will call IR peak in this paper. A Dirac fermion with a small mass $m$ has a sharp peak of Berry curvature in momentum range $|{\bf k}|\sim m$. Indeed, we see that the right peak moves deeper into the IR(larger $n$) when we decrease the mass gap of the Dirac cone. The peak position is approximately $n\sim -\log_2 (|m|)$. (For more discussion about the mass dependence of the IR peak, see Appendix~\ref{appendix: chern number density})

\begin{figure}[htb]
{\includegraphics[scale=1]{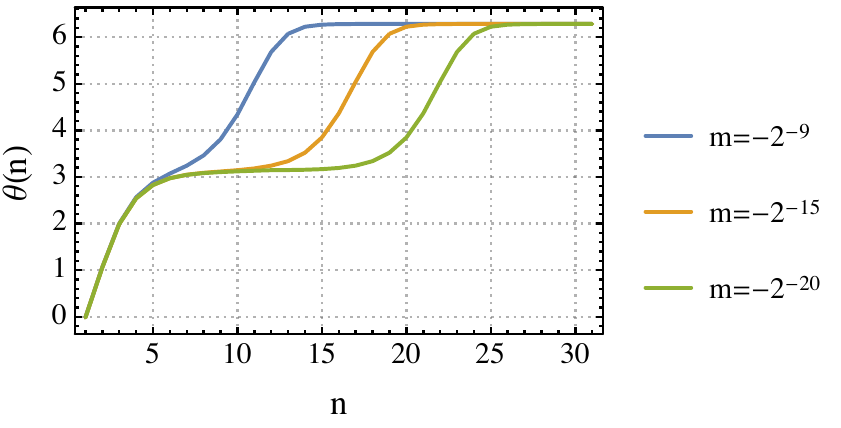}}
\caption{Numerical results of $\theta$ angle for Dirac models with different masses. One can observe that plateaus emerge at $\theta=\pi$ for small masses.}
\label{fig:theta_numerical}
\end{figure}

In the small mass limit $m\ll 1$, each peak contributes half of the Chern number. Physically, this means the Hall conductance $\sigma_{xy}$ of the bulk state concentrates around two surfaces corresponding to the two peak positions, and each surface contributes a Hall conductance $\frac{e^2}{2h}$. This observation immediately tells us that the bulk theory can be viewed as a slab of three-dimensional topological insulator with time reversal symmetry breaking surfaces. The electromagnetic response of a $(3+1)$-d topological insulator is described by the topological term\cite{qi2008topological,essin2009magnetoelectric} (in addition to the ordinary Maxwell terms):
\begin{eqnarray}
\mathcal{L} &=& \frac{\theta}{8\pi^2} \epsilon^{\mu\nu\sigma\tau} \partial_\mu A_\nu \partial_\sigma A_\tau
\label{axion}
\end{eqnarray}
The theta angle
$\theta \in [0,2\pi)$ is determined by the electronic state, and is restricted to $\theta=0$ (trivial insulator) or $\pi$ (topological insulator) for time reversal invariant systems. The topological response theory still applies when there is time reversal symmetry breaking on the surface of TI, which leads to a gapped surface. In this case, the angle $\theta$ deviates away from $0$ or $\pi$ near the boundary. By an integration by parts we obtain $
S=\int d^4 x \mathcal{L}= -\frac{1}{8\pi^2} \int d^4 x  \partial_{\mu}\theta \epsilon^{\mu\nu\sigma\tau } A_\nu \partial_\sigma A_\tau
$, which shows that the gradient of $\theta$ has the physical interpretation of Hall conductance\cite{qi2008topological}. In a system with boundaries in the $z$ direction, we have
\begin{eqnarray}
\sigma_{xy}(z)=\frac{e^2}{2\pi h}\partial_z\theta(z)
\end{eqnarray}
Across each surface of a TI, the phase $\theta$ winds by $\pi$ mod $2\pi$, which leads to a half-integer Hall conductance at each surface. With this understanding, in the EHM case we can obtain the value of $\theta(n)$ as a function of $z$-direction position by integrating over the Chern number density $c(n)$: $\theta(n)=2\pi \sum_{m=1}^nc(m)$. The resulting $\theta(n)$ is shown in Fig.~\ref{fig:theta_numerical}. In the small mass limit, a plateau emerges at $\theta=\pi$, corresponding to the bulk region of TI.

\begin{figure}[h]
\center
\subfloat[Dirac model 1]{\includegraphics[scale=0.3]{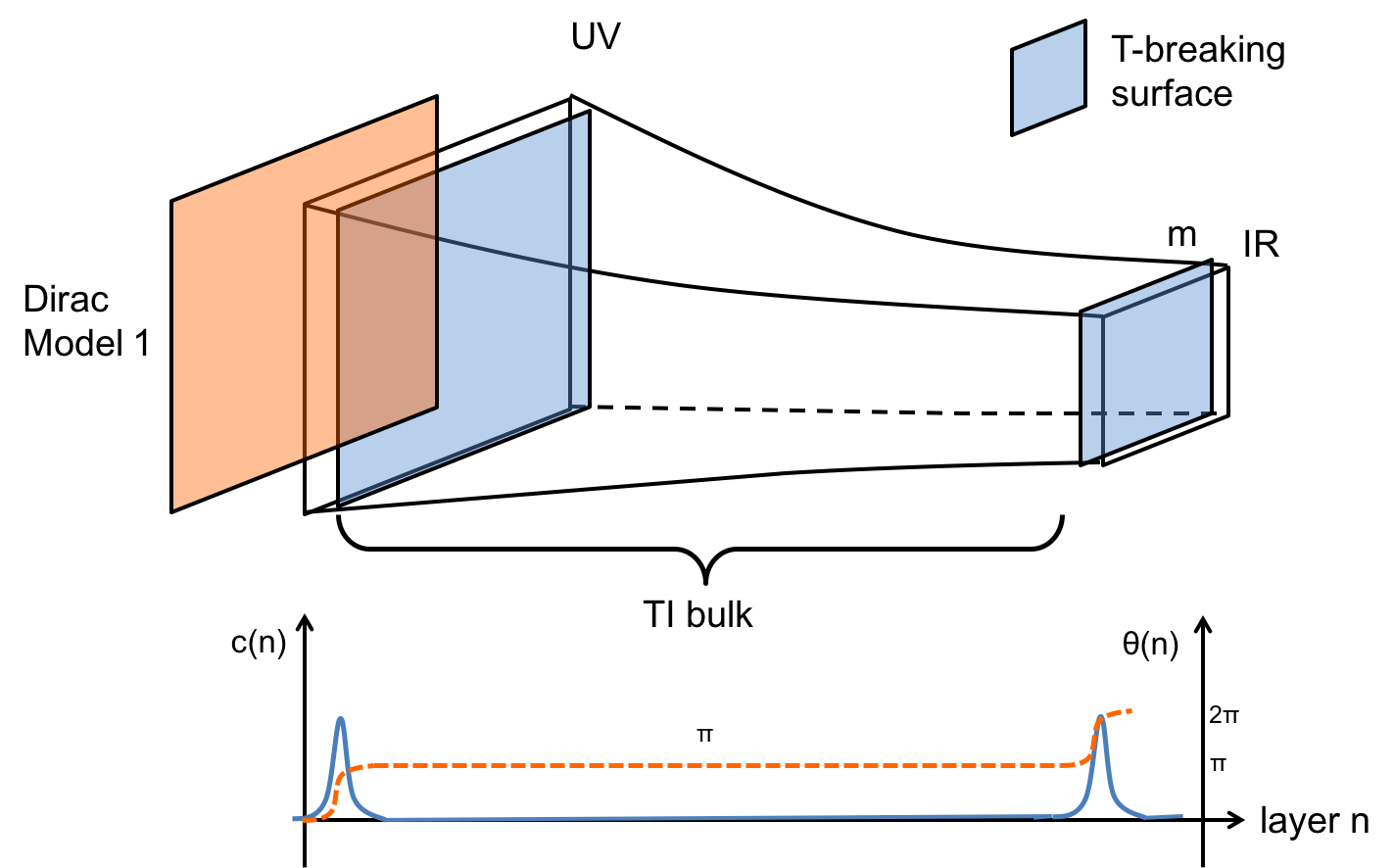}}

\subfloat[Dirac model 2]{\includegraphics[scale=0.3]{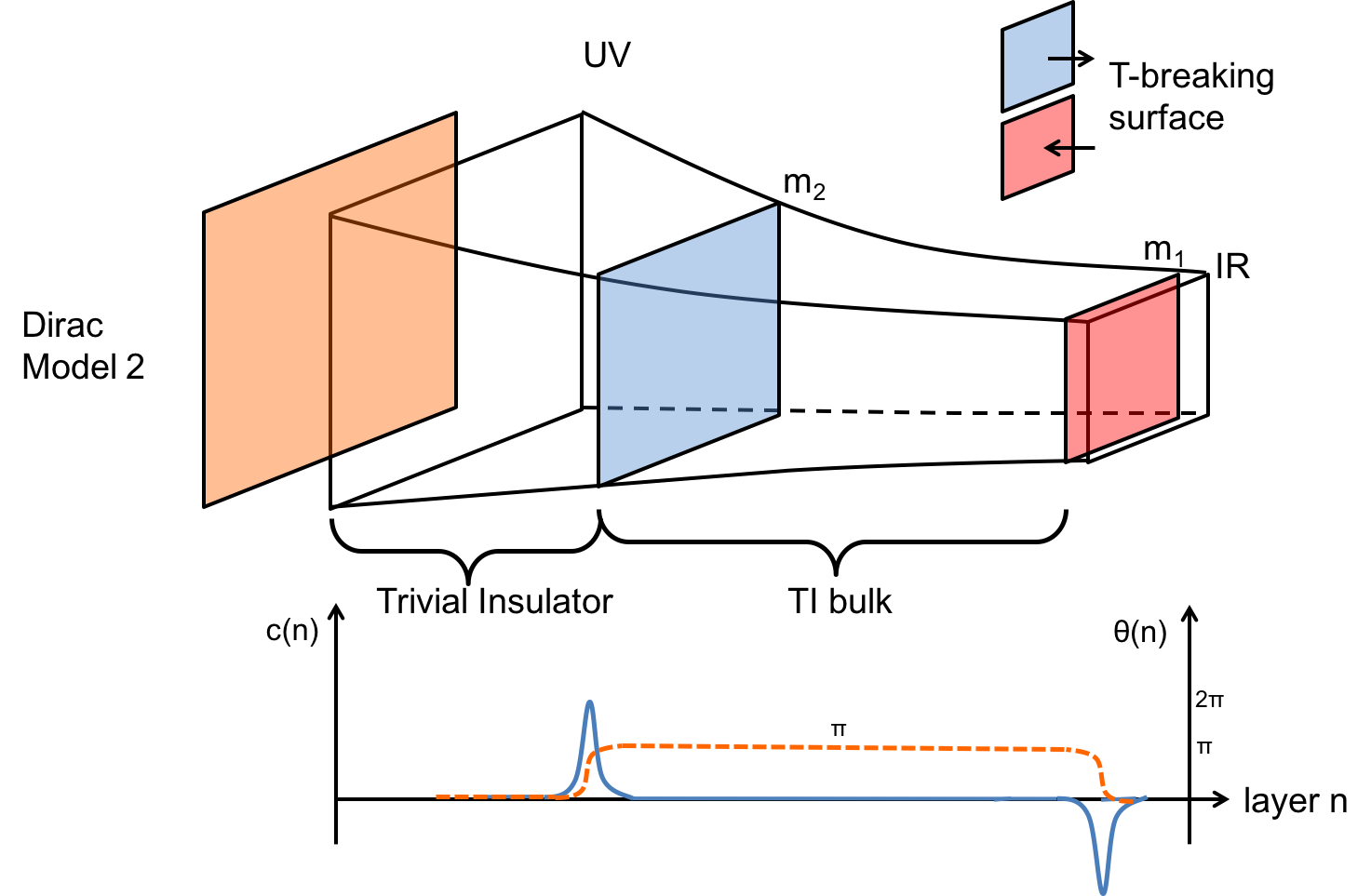}}

\caption{An illustration of holographic topological insulators mapped from Dirac models. (a) The two band model (Eq. \ref{eqn:two band chern insulator model}) which only has one mass scale. (b) The four band model (Eq. \ref{eqn: model 4-band}) which has two mass scales $m_1$ and $m_2$. 
}
\label{fig:bulk holographic TI}
\end{figure}

The holographic topological insulator obtained above has one surface at the scale of the Dirac mass and the other surface at the UV boundary. In general, we can have multiple Dirac cones in a $(2+1)$-d band insulator. For example, consider a simple four band model:
\begin{eqnarray}
\widehat{{H}}&=&\sum_{{\bf k}=(k_x,k_y)} c^\dagger_{\bf k}  h\left({\bf k} \right) c_{\bf k}
\end{eqnarray}
\begin{eqnarray}
\begin{split}
h({\bf k})=& \sin k_x \sigma_x \mathbb{I} + \sin k_y \sigma_y \mathbb{I} \\
+& (m+2-\cos k_x-\cos k_y)\sigma_z\tau_z+ m^\prime \sigma_z \mathbb{I}
\end{split}
\label{eqn: model 4-band}
\end{eqnarray}
Here $\sigma$ and $\tau$ are two sets of Pauli matrices, and the time reversal operator in this basis is $\mathcal{T}=i\sigma_y \tau_x \mathcal{K}$.
$\widehat{{H}}$ describes two decoupled massive Dirac models, with masses $m_1=-m+m^\prime$ and $m_2=m+m^\prime$ ($0<m^\prime<m$). This model has been proposed as the low energy effective description of  Mn doped HgTe quantum well.\cite{liu2008quantum} If $m'=0$, the two are time reversed pairs, and therefore the Berry curvature vanishes everywhere in BZ. In terms of the Chern number density picture introduced in the previous section, $m'=0$ corresponds to the case that two opposite profiles of Chern number density cancel each other. When $m'\neq 0$ the two decoupled Chern insulators are not exactly time reversed pairs, and a pair of opposite peaks of Chern number density emerges in IR at scales $|m_1|$ and $|m_2|$. The Chern number density at UV region remains zero. In the bulk picture (Fig. \ref{fig:bulk holographic TI} (b)), this system corresponds to a TI with opposite T-breaking surfaces in the bulk, with total Hall conductance $\sigma_{xy}$ vanishing.

{
In summary, there are two peaks of Chern number density in the bulk dual of a massive Dirac model, each of which carries Chern number $1/2$. 
One peak (IR) originates from the nontrivial Berry curvature carried by the low momentum region around the Dirac cone, and the other peak (UV) from its lattice regularization near the Brillouin zone boundary. The net Chern number of the two-dimensional system determines the relative sign of these two peaks. In a general model with multiple Dirac cones, there are a sequence of ``branes" in the bulk, each of which carries a Chern number $1/2$. Neighboring bulk regions separated by such a brane are topologically different three-dimensional insulators with opposite $\ZZ_2$ invariants. 
In particular, even a trivial $(2+1)$-d insulator can correspond to a dual theory with TI regions, as long as the net Chern number on all TI surfaces vanish.}

\subsection{Discussions on the bulk geometry}

The multiple mass-scale case, e.g. model~\ref{eqn: model 4-band}, also has an interesting bulk geometry defined via the EHM. The geometry corresponds to such a model is essentially a truncated AdS space consisting of two AdS domains with different curvatures. The approximate metric capturing the asymptotic behavior of the distance is the following:
\begin{equation}
ds^2=\left(\frac{\rho^2}{R^2(\rho)}+1\right)dt^2+\frac{R^2(\rho)}{\rho^2+R^2(\rho)} d\rho^2 +\rho^2 d\Omega^2
\end{equation}
\begin{eqnarray}
d\Omega^2 &=& d \theta^2 + \sin \theta^2 d \phi^2\nonumber \\
R(\rho) &=& \begin{cases} R_2 & \rho_2<\rho < \infty \ ({\rm white})\\
R_1 & \rho_1\leq \rho<\rho_2\ ($light-gray$) \end{cases} \quad \begin{tikzpicture}[scale=0.5,baseline={([yshift=-4pt]current bounding box.center)}]
\draw (0pt,0pt) circle (60pt);
\filldraw[line width=1pt,draw=black,light-gray] (0pt,0pt) circle (50pt);
\filldraw[gray](0pt,0pt) circle (20pt);
\draw[line width=0.8pt,>=stealth,->](0pt,0pt)--(16pt,12pt) node[right]{$\rho_1$};
\draw[line width=0.8pt,>=stealth,->](0pt,0pt)--(24pt,43pt) node[black,left]{$\rho_2$};
\end{tikzpicture}\nonumber
\end{eqnarray}

More precisely, we determine the geometry from the physical mutual information according to the general principle of EHM, see Ref.~\onlinecite{qi2013exact}. In a massive theory, the mutual information between two distant point has the form\cite{lee2015}$I(d) \sim {C}/{d^6}$ asymptotically ($d< {1}/{m}$), and has a exponential decay $I(d) \sim I_0 e^{-d/\xi}$ when $d>{1}/{m}$. In a theory with multiple mass scales, there are multiple cross over scales where the correlation behavior changes. They corresponds to ``domain walls" in the $z$ direction in the bulk geometry. 
More concretely, in the model Eq.\ \eqref{eqn: model 4-band}, the two Dirac cones decouple, so we have $I(d)=I_1(d)+I_2(d)$. Therefore for $d<{1}/{m_2}$, we have $I(d) \sim {2C}/{d^6}$, and for ${1}/{m_2}	<d<{1}/{m_1}$ we have $I(d) \sim \frac{C}{d^6}$. For $d>{1}/{m_1}$ $I(d)$ decays exponentially. Comparing this with the asymptotic behavior in AdS space\cite{qi2013exact,lee2015} (by assuming the mutual information decays exponentially with AdS distance)
\begin{eqnarray}
I(d)=I_0 \exp \left(-\frac{R}{\xi} \log \left(\frac{d}{R} \right) \right),
\end{eqnarray}
one obtains $\frac{R_1}{\xi_1}=\frac{R_2}{\xi_2}=6$ and $R_2=\sqrt[6]{2} R_1$. 

It is worth mentioning that the analysis here is consistent with intuition from the AdS/CFT correspondence in the large $N$ limit, where the AdS radius $R$ of the semiclassical geometry is determined by the central charge of the boundary CFT. Here, the central charge of the boundary model Eq.\ \eqref{eqn: model 4-band} decreases $2$ to $1$ and then to $0$ in renormalization group flow upon increase of the length scale, corresponding to the $z$ dependence of AdS radius. Such a decrease of central charge is, of course, consistent with the $c$-theorem.\cite{zamolodchikov1986,freedman1999renormalization,girardello1999novel,myers2010seeing} 

\section{Identifying the bulk theory (II): entanglement spectrum}\label{sec:bulktheory2}

A defining feature of topological insulators is the presence of gapless surface states on time-reversal invariant surfaces. As was discussed above, the surfaces we naturally have in the 3 dimensional EHM bulk are not time reversal symmetric, so there is no gapless surface at hand. However, there is another kind of surface states that characterize the topological nature of the TI bulk, which are those in the entanglement spectrum.
\cite{2006PhRvB..73x5115R,turner2010entanglement,fidkowski2010entanglement,pollmann2010entanglement}

\begin{figure}[htb]
\includegraphics[scale=0.3]{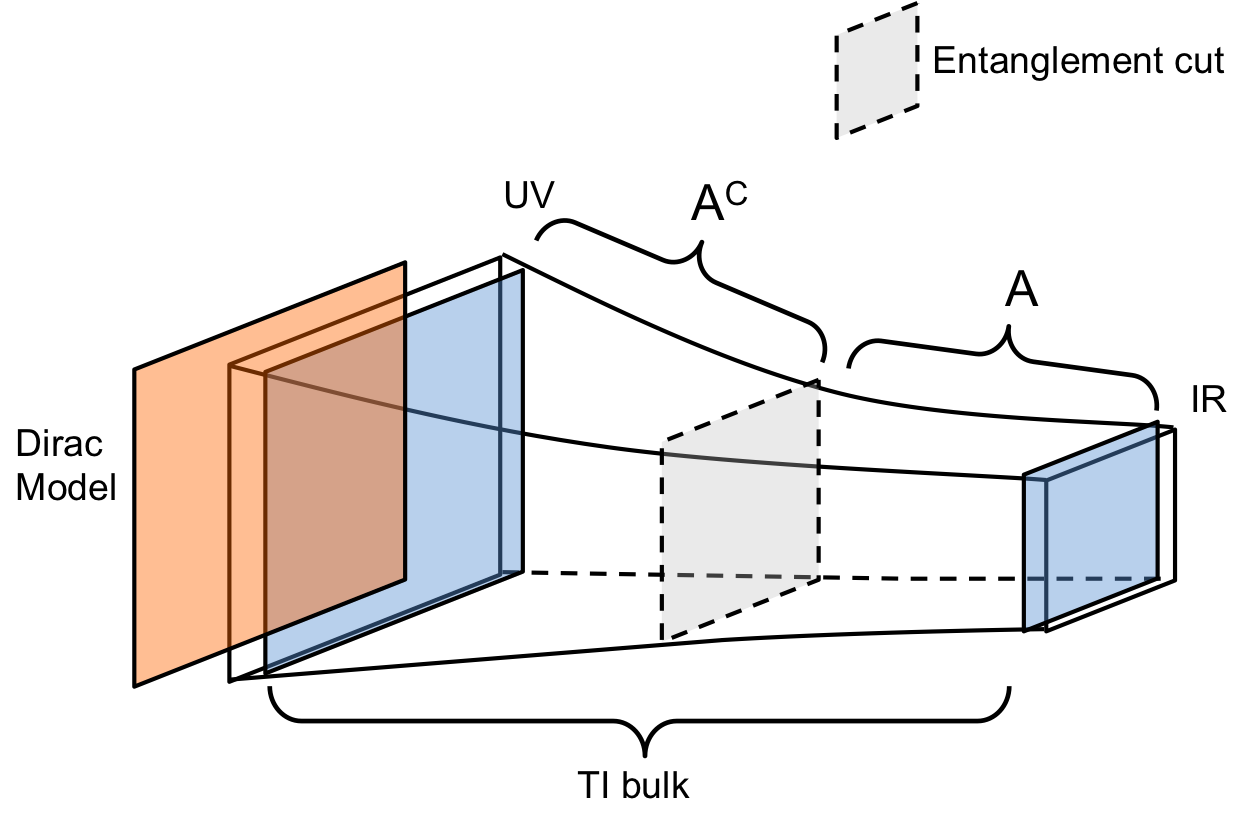}
\caption{An entanglement cut in the bulk of the holographic TI (dashed rectangle) divides the TI into two parts ${\rm A}$ and ${\rm A^C}$. The reduced density matrix of $A$ is obtained by a partial trace over region ${\rm A^C}$.}
\label{fig:EC1}
\end{figure}

In order to study the gapless surface states in entanglement spectrum, we introduce an entanglement partition in the interior of TI region, where time reversal symmetry is preserved. 
More explicitly, as shown in Fig.~(\ref{fig:EC1}), we divide the whole system into two parts A and ${\rm A^C}$. The density matrix of the state in the whole system is denoted as $\rho=|\psi\rangle \langle \psi |$, with $|\psi\rangle$ the ground state. The reduced density matrix $\rho_A$ and entanglement Hamiltonian~\cite{peschel2002calculation, lee2014,lee2015free} $H_A$ for subsystem A are defined by the partial trace:
\begin{eqnarray}
\rho_A : = \Tr_{\rm A^C} \left(|\psi\rangle \langle \psi | \right) =: e^{-H_A}
\end{eqnarray}
The entanglement spectrum is defined as the eigenvalue spectrum of $H_A$.

We choose the entanglement cut to be a constant $z$ plane, such that we have a residual translation symmetry in the bulk layers. Therefore we are still able to talk about the entanglement spectrum as a function of momentum ${\bf q}$ in the folded Brillouin zone. Furthermore, for non-interacting fermions, $H_A$ is also quadratic\cite{peschel2002calculation}, which can be written as $H_A=\sum_{\bf q}a_{\bf q}^\dagger h_A({\bf q})a_{\bf q}$. Here $h_A({\bf q})$ is a single-particle entanglement Hamiltonian with multiple bands, with $a_{\bf q}$ the corresponding annihilation operators. $h_A({\bf q})$ can be determined by the two-point functions
$C_{\rm A}({\bf q}):=\operatorname{P}_{\rm A} \langle a^{\dagger}({\bf q}) a({\bf q}) \rangle  \operatorname{P}_{\rm A}$, which is the two-point function of the ground state truncated\cite{peschel2002calculation} to region $A$:
\begin{equation}
h_{\rm A}({\bf q})=\log \left[C_{\rm A}({\bf q})^{-1}-\mathbb{I}\right]
\end{equation}
With this, we can calculate the entanglement spectrum via the explicit EHM transformation and the boundary single particle hamiltonian\eqref{eqn:two band chern insulator model}.

\begin{figure}[h]
\center
\subfloat[Entanglement spectrum for cut-position $n=2$, $14$ and $30$]{
\includegraphics[scale=0.33]{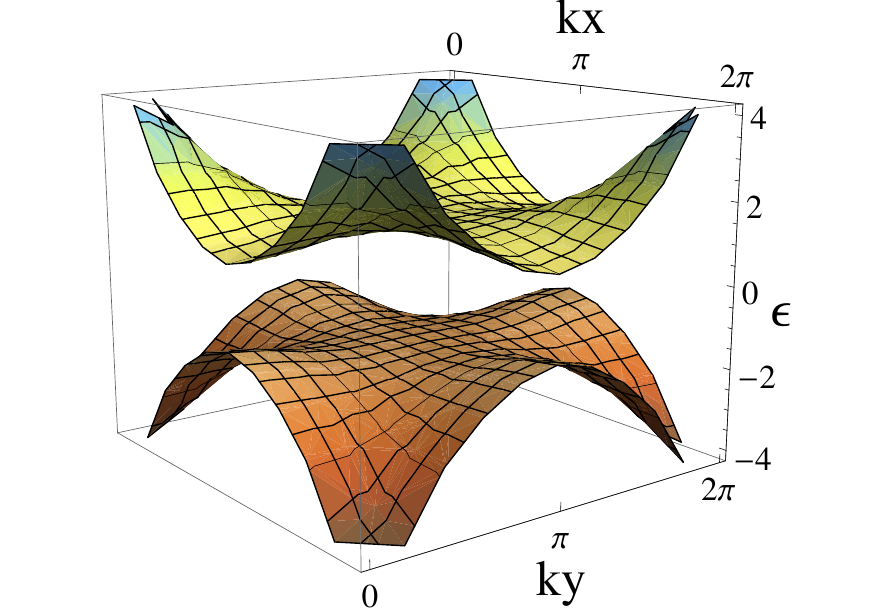}\includegraphics[scale=0.33]{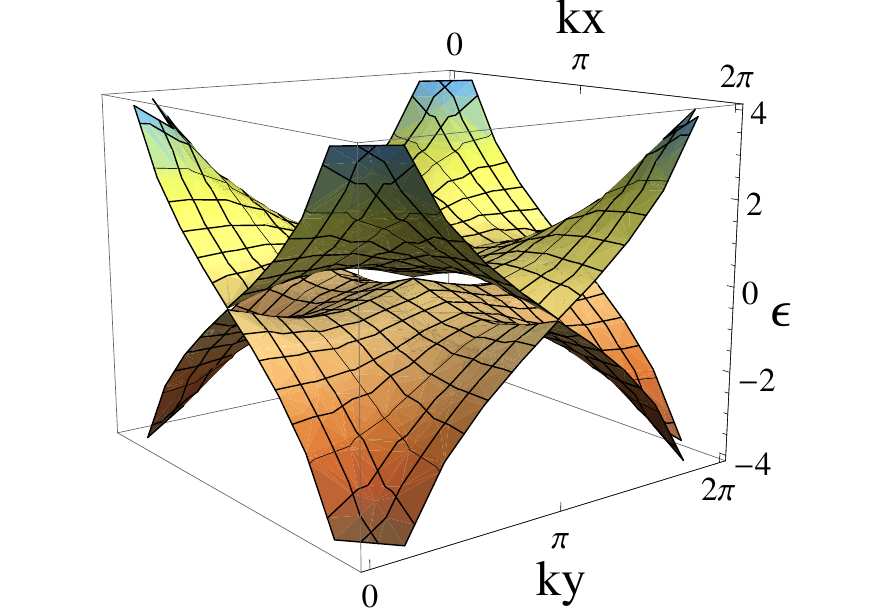}\includegraphics[scale=0.33]{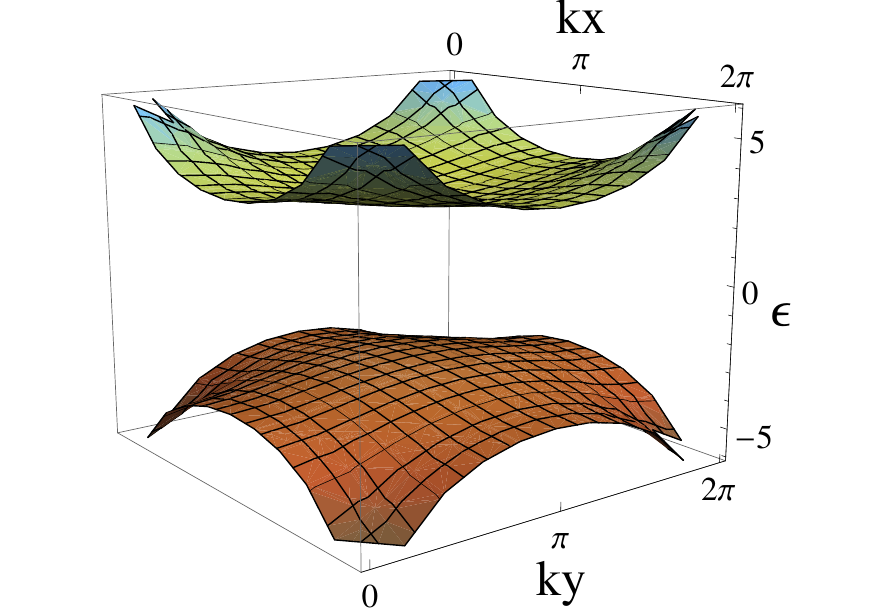}}

\subfloat[Minimal gap of the correlator as function of $n$]{
\includegraphics[scale=0.25]{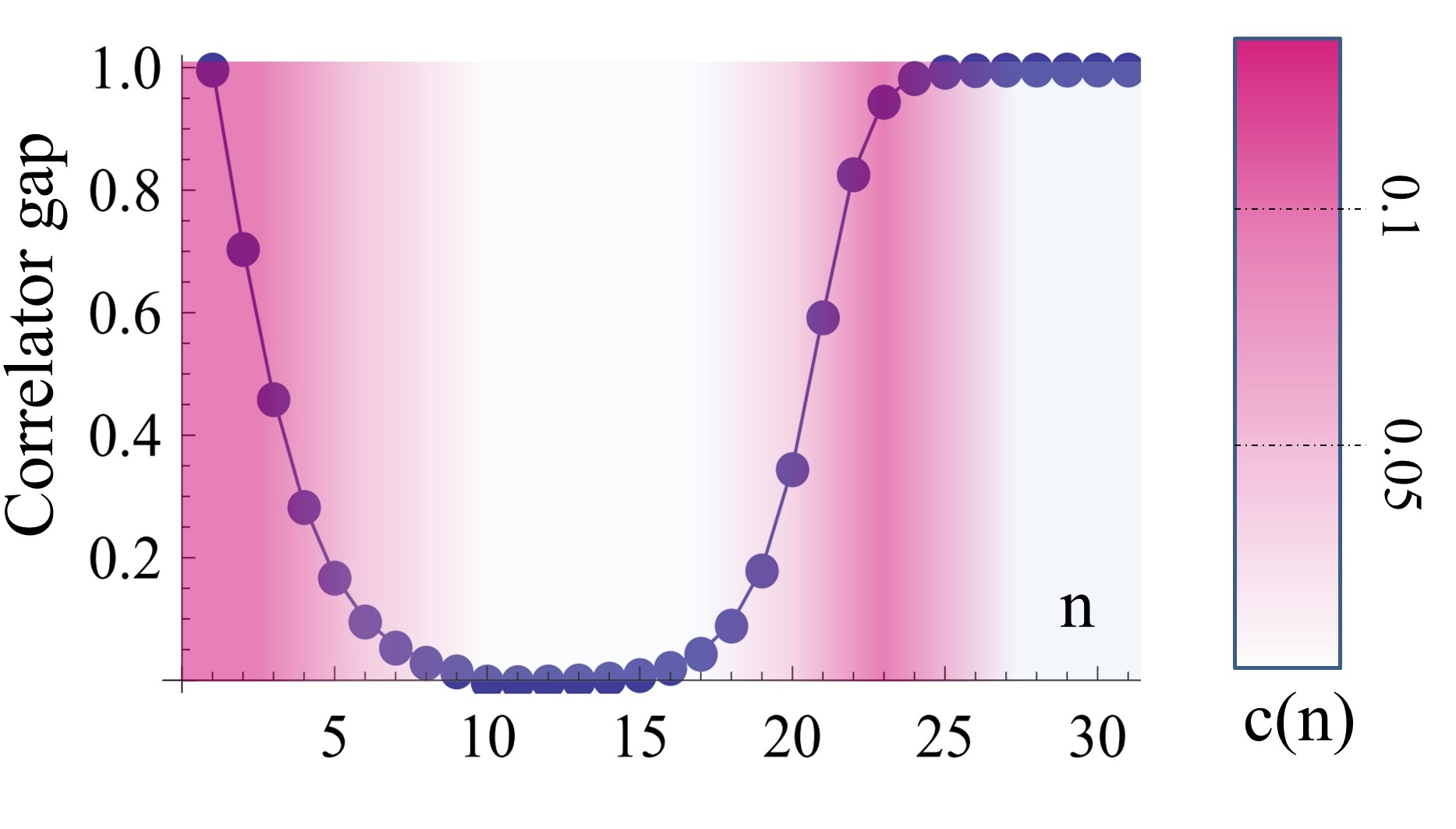}
}
\caption{(a) The low energy entanglement spectra $\epsilon({\bf q})$ for different positions of entanglement cuts at layers $n=2,14$ and $30$ of the EHM bulk (Eq.\ \eqref{eqn:two band chern insulator model}), with the small mass set to $m=-2^{-20}$. The spectra are gapped for cuts in deep UV ($n=2$) and deep IR ($n=30$) regimes, but is gapless when the cut is made in the bulk TI region ($n=14$). The entanglement gap closes at three massless Dirac points $(\pi,0)$,$(0,\pi)$ and $(\pi,\pi)$ in the reduced Brillouin zone. (b) Plot of the minimal gap of the correlator $C_A({\bf q})$ spectrum. 
The background color density represents the Chern number density $c(n)$ profile. Indeed, the entanglement gap vanishes only in the TI bulk region, where $c(n)$ is exponentially small between its two peaks in the UV and the IR. }
\label{fig:ES}
\end{figure}


The numerical results for low energy entanglement spectrum are shown in Fig.~(\ref{fig:ES}a) for different position of the entanglement cut. The mass $m=-2^{-20}$ is chosen to be very small, so that the time reversal breaking surfaces in the UV and IR are well-separated. In the middle plot ($n=14$) depicting the regime between the UV and IR $c(n)$ peaks, where the time reversal breaking $c(n)$ is almost vanishing, the gap closes at three momentum points $(\pi,0)$, $(0,\pi)$ and $(\pi,\pi)$, each with a linear dispersion. The odd number of Dirac cones in the entanglement spectrum is consistent with the nontrivial $\mathbb{Z}_2$ topological nature of the TI bulk.\cite{turner2010entanglement,fidkowski2010entanglement} In Fig.~(\ref{fig:ES}b), we also demonstrate a systematic scan of the minimal gap in the eigenvalue spectrum of $C_A({\bf q})$, which directly determines the gap of entanglement spectrum. The background is colored according to the Chern number density, which indicates the location of physical surfaces with time-reversal breaking. Indeed, the entanglement gap only closes in the $\mathbb{Z}_2$ nontrivial regime between the two $c(n)$ peaks.

From the point of view of the $(2+1)$-d quantum Hall state, the entanglement spectrum with a $z$-direction partition we studied here provides a new measure of the intrinsic coupling between UV and IR degrees of freedom in this system. In the ordinary renormalization group approach, the low energy degrees of freedom is assumed to be in a pure state---the ground state of the low energy effective field theory, after the high energy degrees of freedom are integrated out. In contrast, in the EHM approach, UV and IR degrees of freedom are both preserved in the Hilbert space, so that the quantum entanglement between them can be studied. 


\section{Conclusion and discussions}\label{sec:conclusion}

In this paper, we used a $(2+1)$-d version of the EHM to generate a quantum state in $(3+1)$-d and studied the topological properties of the bulk. By defining the Chern number density, which corresponds to the Hall conductance of a single layer in the $(3+1)$-d system, we identified the quantum state we generated as a time-reversal invariant topological insulator with time-reversal breaking surfaces. The locations of the TI surfaces in the emergent $z$ direction are determined by the correlation length scales in the $(2+1)$-d model. We further studied the entanglement properties of the state to verify this claim. By an entanglement cut in the $z$ direction, we obtain a gapless entanglement spectrum, which is a new signature of the intrinsic entanglement between UV and IR degrees of freedom in the $(2+1)$-d state.

The gapless UV-IR entanglement spectrum also provides a new entanglement measure of quantum anomalies. If we take the limit $m\rightarrow 0$ in the Dirac model, the bulk dual theory is still a TI, with its IR boundary pushed to the center of the AdS disk. The entanglement spectrum between UV and IR is still gapless as long as the entanglement cut is at a length scale that's much larger than the UV cutoff scale. In this limit, the IR effective theory of the $(2+1)$-d system is a massless Dirac fermion, which is known to have parity anomaly. 
\cite{Redlich:1983kn, Redlich:1983dv}
In condensed matter theory language, the parity anomaly means the single Dirac cone cannot be defined in a lattice $(2+1)$-d theory without breaking time-reversal symmetry. The UV regularization always breaks time-reversal symmetry and contributes Chern number $1/2$. The gapless UV-IR entanglement spectrum in the bulk dual theory is an entanglement measure of the intrinsic coupling between IR and UV degrees of freedom in the $(2+1)$-d state, as a consequence of the parity anomaly.
Recently, analogous analyses of the entanglement between the UV and IR degrees of freedom have been employed in studying quantum criticality in a topological state of a free fermion system, see Ref.~\onlinecite{hsieh2014bulk,vijay2014entanglement,lai2015entanglement}, where either a symmetric partition or a random partition separates the integer Chern number into two half-integers, thereby producing nontrivial entanglement.


The EHM approach can be generalized straightforwardly to higher dimensional topological states. The correspondence between different topological states with dimension $d$ and $d+1$ is generic, and can be seen explicitly by applying EHM to lattice Dirac models. In general, the EHM approach provides a new description of topological properties of a fermion model in geometry and entanglement measures. More detailed investigation of the higher dimension cases will be reserved for future works.

\noindent{\bf Acknowledgement}

We would like to acknowledge helpful conversation with Wenbo Fu, Chao-Ming Jian, Ronny Thomale and Bo Yang. 
This work is supported by
the NSF under Grants 
No. DMR-1151786 (YG and XLQ), 
No. DMR-1455296  (X.W.  and S.R.), 
the  Alfred  P.  Sloan  foundation (SR), 
the Brain Korea 21 PLUS Project of Korea Government (GYC), 
as well as by the David and Lucile Packard Foundation (XLQ).
We are grateful to the KITP Program
``Entanglement in Strongly-Correlated Quantum  Matter" 
(Apr 6 - Jul 2, 2015). 
This research was supported in part by the National Science Foundation under Grant No. NSF PHY11-25915.

\appendix

\section{Detailed derivation of the EHM transformation in momentum space}
\label{appendix: momentum space}

In this appendix, we derive the momentum space EHM transformation according to its definition in the main text (Eq. \ref{eqn: real space EHM}), which is written in real space as:
\begin{eqnarray}
V_n&=&V^y_n V^x_n\\
V^x_n:
\left( \begin{tabular}{c}
$\tilde{a}_{i,j,n}$ \\
$b^x_{i,j,n}$
\end{tabular}
\right) &=& \frac{1}{
\sqrt{2}}
\left(
\begin{tabular}{c c}
$1$ & $1$ \\
$1$ & $-1$
\end{tabular}
\right)
\left( \begin{tabular}{c}
$a_{2i-1,j,n-1}$ \\
$a_{2i,j,n-1}$
\end{tabular}
\right) \nonumber \\
V^y_n:
\left( \begin{tabular}{c}
$a_{i,j,n}$ \\
$b^{y}_{i,j,n}$
\end{tabular}
\right) &=& \frac{1}{
\sqrt{2}}
\left(
\begin{tabular}{c c}
$1$ & $1$ \\
$1$ & $-1$
\end{tabular}
\right)
\left( \begin{tabular}{c}
$\tilde{a}_{i,2j-1,n}$ \\
$\tilde{a}_{i,2j,n}$
\end{tabular}
\right)  \nonumber
\end{eqnarray}
$\tilde{a}$ stands for an auxiliary IR mode, which is the input of next step $V^y$. In momentum space, we keep the same notation in naming the modes, but with subscripts being momentum: e.g. $a_{{\bf q},n}$. For convenience, we rescale the folded Brillouin zone after each step to the standard one $[-\pi,\pi)\times [-\pi,\pi)$ and shift by $(\pi,\pi)$ for the purpose in this section, i.e. each momentum ${\bf q}=(q_x,q_y)\in [0,2\pi)\times [0,2\pi)$. The $x$-direction transformation $V_n^x$ in momentum space is:
\begin{equation}
\left(
\begin{tabular}{c}
$\tilde{a}_{q_x,q_y,n+1}$\\
$b^x_{q_x,q_y,n+1}$
\end{tabular}
\right)=
\left(
\begin{tabular}{cc}
$C(q_x/2)$ & $D(q_x/2)$ \\
$D(q_x/2)$ & $C(q_x/2)$
\end{tabular}
\right)\left(
\begin{tabular}{c}
$a_{q_x/2,q_y,n}$ \\
$a_{q_x/2+\pi,q_y,n}$
\end{tabular}
\right)
\end{equation}
where $C(q),D(q)$ are, for the Haar wavelet (details in \cite{lee2015}):
\begin{eqnarray}
C(q):= \frac{1+e^{-iq}}{2} \qquad D(q):=\ \frac{1-e^{-iq}}{2}
\end{eqnarray}
which also satisfy the useful properties $
C(q)D(q)=D(2q)/2$, $|C|^2+|D|^2=1$ and $C(q+\pi)=D(q)$. Similarly, the second step action on the y-direction has the expression:
\begin{equation}
\left(
\begin{tabular}{c}
$a_{q_x,q_y,n+1}$\\
$b^y_{q_x,q_y,n+1}$
\end{tabular}
\right)=
\left(
\begin{tabular}{cc}
$C(q_y/2)$ & $D(q_y/2)$ \\
$D(q_y/2)$ & $C(q_y/2)$
\end{tabular}
\right)\left(
\begin{tabular}{c}
$\tilde{a}_{q_x,q_y/2,n+1}$ \\
$\tilde{a}_{q_x,q_y/2+\pi,n+1}$
\end{tabular}
\right)
\end{equation}
Applying the above transformations recursively, we get the formula for basis transformation of arbitrary layer index $n$ in terms of boundary bases $a_{{\bf k},0}$
\begin{eqnarray}
a_{{\bf q},n}&=& \sum_S Y({\bf k},n) a_{{\bf k},0}  \\
b^x_{{\bf q},n}&=& \sum_{S} D(2^{n-1}k_x) Y({\bf k},n-1)a_{{\bf k},0} \\
b^y_{{\bf q},n}&=&\sum_{S} D(2^{n-1}k_y)C(2^{n-1}k_x) 
Y({\bf k},n-1) a_{{\bf k},0}
\end{eqnarray}
where $S$ is a set of points ${\bf k}=(k_x,k_y)$ (this set depends on ${\bf q}=(q_x,q_y)$ and $n$) in Brillouin zone: $S=S({\bf q},n):=\lbrace{\bf k}=(k_x,k_y)| k_x={q_x+2\pi m}/{2^n}, k_y={q_y+2\pi l}/{2^n}; 0 \leq l,m\leq 2^n-1 \in \mathbb{Z} \rbrace$. $Y({\bf k},n)$ is a complex valued function:
\begin{equation}
Y({\bf k},n):=\prod_{i=0}^{n-1}C(2^i k_x)  C(2^i k_y)
=\frac{D(2^n k_x) D(2^n k_y)}{2^{2n} D(k_x)D(k_y)}
\end{equation}
Here, the two UV modes generated by the $x$-direction EHM are collectively called $b^x$. Hence the 3 real space UV modes in the main text corresponds to the two $b^x$'s and $b^y$.

\begin{figure}[htb]
{\includegraphics[scale=0.89]{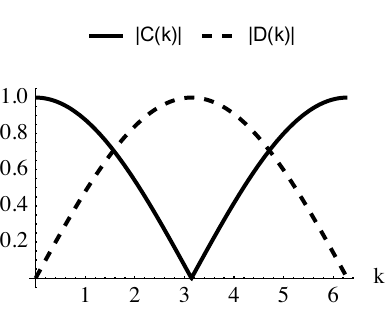}}
{\includegraphics[scale=0.89]{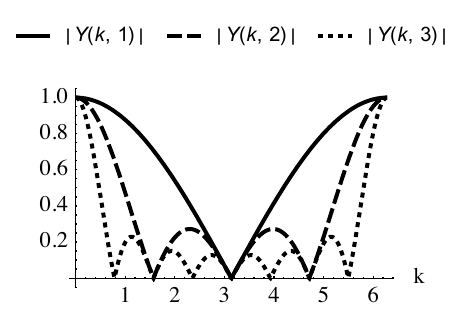}}
\caption{Plot for functions $C(k)$, $D(k)$ and $Y(k,n)$ for $k_y=0$. The label $k$ in figure represents $k_x$ in our EHM. Shown are the plots for layers $n=1,2,3$.}
\label{fig:Yplot}
\end{figure}

\section{Details for the derivation and approximations of the Berry curvature and Chern number density for the Dirac model}
\label{appendix: chern number density}

With the basis transformation in last section, it is straightforward to deduce the projected Chern number to $n$ layer's IR mode $a_n$:
\begin{eqnarray}
c^{\rm IR}(n) = \frac{1}{2\pi} \int_{BZ} dk^2 |Y({\bf k},n)|^2 f_0({\bf k})
\label{eqn:IR chern number density}
\end{eqnarray}
The Chern number density defined in main text is related to $c^{\rm IR}$ by
\begin{eqnarray}
c(n)=c^{IR}({n})-c^{IR}({n+1})
\end{eqnarray}
In Fig 8, the Chern number density for small $m$ is exactly numerically computed up to layers $n=9$. Beyond that, an approximation of the Berry curvature is necessary to avoid the rapid oscillation of $|Y(\bold k,n)|^2$. Below, we shall demonstrate how this can be done to reveal the ultimate separation of the ``IR peak'' from the UV contributions.

An approximated form of Berry curvature for the $|{\bf k}|\ll 1$ region is used to extract the generic behavior for the IR peak for a small mass $m\ll 1$. Near the $\Gamma$ point,
\begin{eqnarray}
f_0({\bf k})= \frac{\vec d\cdot (\partial_{k_x} \vec d \times \partial_{k_y} \vec d)}{2 |d|^3} \simeq \frac{m } {2(m^2+k_x^2+k_y^2)^{3/2}}
\label{eqn: berry curvature: near gamma approximation}
\end{eqnarray}
where $\vec d=\vec{d}({\bf k})$ is the d-vector in the model (Eq.\ \ref{eqn:two band chern insulator model}). It is also useful to mention that $\int dk^2 \frac{m } {2(m^2+k_x^2+k_y^2)^{3/2}}=\pi$, which further justify our replacement of the IR region of $f_0({\bf k})$ by such an `idealized' Berry curvature. Moreover, we can also approximate the weight function in Eq.\ \eqref{eqn:IR chern number density} for the $|{\bf k}|\ll 1$ region:
\begin{eqnarray}
 |Y({\bf k},n)|^2&=&2^{-4n} \left( \frac{\sin \left( 2^{n-1} k_x\right) \sin \left( 2^{n-1} k_y\right)}{\sin \left( 2^{-1} k_x\right) \sin \left( 2^{-1} k_y\right)} \right)^2\nonumber  \\
 &\simeq & 2^{-4n} \left( \frac{4\sin \left( 2^{n-1} k_x\right) \sin \left( 2^{n-1} k_y\right)}{k_x k_y} \right)^2
\end{eqnarray}
Now we define a function $I^{\operatorname{IR}}(\alpha,m)$, where $\alpha=2^n$, to capture the behavior of $c^{\operatorname{IR}}(n)$ for the IR side $n\gg 1$ (we also made the dependence on mass explicit for later convenience):
\begin{eqnarray}
I^{\operatorname{IR}}(\alpha,m) := \frac{1}{2\pi} \int &dk^2& \left( \frac{4\sin \left( \alpha k_x/2\right) \sin \left( \alpha k_y/2\right)}{k_x k_y} \right)^2\nonumber \\&& \times\frac{m } {2\alpha^4 (m^2+k_x^2+k_y^2)^{3/2}}
\end{eqnarray}
Here the integration region has been extended from the Brillouin zone to whole plane, which is permissible giving the rapidly decaying integration kernel. The Chern number density for the IR region is related to $I^{\operatorname{IR}}(\alpha,m)$ by:
\begin{eqnarray}
{\rm IR}: \quad c(n) \simeq I^{\operatorname{IR}}(2^{n},m)-I^{\operatorname{IR}}(2^{n+1},m)
\end{eqnarray}
Regarding $\alpha$ as a continuous variable, we have the symmetry for $I^{\operatorname{IR}}(\alpha,m)$:
\begin{eqnarray}
I^{\operatorname{IR}}(\alpha,m)=I^{\operatorname{IR}}(\lambda \alpha, \lambda^{-1}m)
\end{eqnarray}
In other words, the IR peak of the Chern number density will be shifted to the right by one layer while keeping the same shape if we reduce the mass by half. Therefore, we can conclude that the bulk TI in the main text is well defined in the zero mass limit, where the IR peak of Chern number density is pushed to $n\rightarrow \infty$.

\section{Single particle projector $C_n$ and symmetry discussion about gap closing point}

For a free fermion model, we can compute the entanglement Hamiltonian via the single particle correlator, see Ref.~\onlinecite{peschel2002calculation}. For the purpose of this paper, we need to calculate the entanglement Hamiltonian for the subsystem of the EHM bulk with layer index $\geq n$. Such subsystem is equivalent to the IR mode at layer n, whose basis is:
\begin{equation}
a_{{\bf q},n}= \sum_{S} Y({\bf k},n) a_{{\bf k},0}
\end{equation}
Therefore, the single particle correlator is related to boundary single particle correlator $G({\bf k})= \langle a^\dagger_{{\bf k},0} a_{{\bf k},0} \rangle  $ via:
\begin{eqnarray}
C_n({\bf q})=\langle a_{{\bf q},n}^\dagger a_{{\bf q},n} \rangle
= \sum_{S} |Y({\bf k},n)|^2 G ({\bf k})
\end{eqnarray}
As a reminder, the set $S=S({\bf q},n):=\lbrace{\bf k}=(k_x,k_y)| k_x={q_x+2\pi m}/{2^n}, k_y={q_y+2\pi l}/{2^n}; 0 \leq l,m\leq 2^n-1 \in \mathbb{Z} \rbrace$. Physically speaking, similar to real space, the k-space form of the EHM transformation is a ``sampling'' over the Brillouin zone of the boundary band insulator. The weight of such sampling is given by function $Y$ for IR mode $a$. The size of the sample set $S$ grows exponentially with layer index $n$. As shown in Fig.~\ref{fig:Yplot}, the function $Y$ is localized near the $\Gamma$ point for large $n$, so it will be permissible to keep a smaller sampling set $S' \subset S$, where $S'$ only contains $\chi$ points that are closest to $\Gamma$, $\chi$ an integer determined by the required accuracy. Thus we can approximate the correlator:
\begin{eqnarray}
C_n({\bf q})=\langle a_{{\bf q},n}^\dagger a_{{\bf q},n} \rangle
\simeq \sum_{S' \subset S} |Y({\bf k},n)|^2 G({\bf k})
\label{eqn: correlator}
\end{eqnarray}
The point of such approximation is that the function $G(\bf k)$ has a simple geometrical picture near $\Gamma$. More explicitly, consider the two band model Eq. \ref{eqn:two band chern insulator model}:\begin{eqnarray}
G ({\bf k}) = \frac{\operatorname{I}-\hat d({\bf k})\cdot \vec\sigma}{2}, \quad
\hat d({\bf k})= \frac{\vec{d}({\bf k})}{|\vec{d}({\bf k})|},
\end{eqnarray}
$\vec{d}({\bf k})=\left( \sin k_x, \sin k_y,m+2-\cos k_x-\cos k_y\right)$ the d-vector for the two band model. There are two asymptotic limits for $\hat{d}({\bf k})$ near $\Gamma$: (1) $k\ll m \ll 1$, $\vec{d}({\bf k}) \simeq \left(0, 0, m \right)$, therefore $\hat{d}({\bf k})\simeq \left( 0, 0, 1 \right)$ is a constant vector. (2) $m\ll k\ll 1$ then  $\vec{d}({\bf k})=\left( k_x,  k_y, 0\right)$ and $\hat{d}({\bf k})\simeq (k_x,k_y,0)/|{\bf k}|$ is an in-plane pointer. Notice that the density of points in sampling set $S(q,n)$ is determined by $n$: $\Delta k =  2\pi /2^n$, when layer index $n\gg \log_2 (|m|)$, the sampling set $S'\subset S$ is in the limit (1), where they pick up constant $\hat{d}\simeq (0,0,1)$, therefore, the correlator $C_n$ is gapped, which is the case for deep IR: $n\gg \log_2(|m|)$. When layer index lives in the ``bulk of TI'' region, where $m \ll \Delta k \ll 1$, the sampling set $S'$ is in limit (2), which is most interesting case. At a generic point, the spectrum is gapped, but there are in total 3 special symmetric points: ${\bf} q=(\pi,0),(0,\pi)$ and $(\pi,\pi)$, where the elements in sampling set $S$ are paired under reflection symmetry: ${\bf k} \rightarrow -{ \bf k}$. Notice, the norm square $|Y(k)|^2$ is even under this reflection symmetry, but $\hat{d}({\bf k})\simeq (k_x,k_y,0)/|{\bf k}|$ is odd under reflection. Therefore,
\begin{eqnarray}
C_n({\bf q})&\simeq &  \sum_{ S} |Y({\bf k},n)|^2 \frac{\mathbb{I}-\hat d({\bf k})\cdot \vec\sigma}{2}\nonumber \\
&\simeq& \frac1{2} \sum_{ S} |Y({\bf k},n)|^2 \mathbb{I}
\end{eqnarray}
which has degenerate eigenvalues, and thus represent gap closure in entanglement spectrum.

In fact, the above symmetry argument about the gap closing points can be generalized to higher dimensional cases, where, we will expect $2^d-1$ gap closing points for a $d$-dimensional Dirac model with a proper generalization of $d$-dimensional EHM.

\bibliography{refs}

\end{document}